\documentclass[prb,aps,amsmath,amssymb,amsfonts,twocolumn,superscriptaddress]{revtex4-1}
\usepackage{CJK} 
\usepackage{graphicx} 
\usepackage{multirow} 
\usepackage{bm} 
\usepackage[colorlinks=true,linkcolor=blue,anchorcolor=red,citecolor=blue, urlcolor=blue]{hyperref}

\begin{document} 
\begin{CJK*}{UTF8}{} 

\title{Quantum Transport in Weyl Semimetal Thin Films in the Presence of Spin-Orbit Coupled Impurities}
\author{Weizhe Edward Liu  (\CJKfamily{gbsn}刘玮哲)}
\affiliation{School of Physics and Australian Research Council Centre of Excellence in Low-Energy Electronics Technologies, UNSW Node, The University of New South Wales, Sydney 2052, Australia}
\author{Ewelina M. Hankiewicz}
\affiliation{Institute for Theoretical Physics and Astrophysics, W\"urzburg University, Am Hubland, 97074 W\"urzburg, Germany}
\author{Dimitrie Culcer}
\email{d.culcer@unsw.edu.au}
\affiliation{School of Physics and Australian Research Council Centre of Excellence in Low-Energy Electronics Technologies, UNSW Node, The University of New South Wales, Sydney 2052, Australia}
\begin{abstract}
Topological semimetals have been at the forefront of experimental and theoretical attention in condensed matter physics. Among these, recently discovered Weyl semimetals have a dispersion described by a three-dimensional Dirac cone, which is at the root of exotic physics such as the chiral anomaly in magnetotransport. In a time reversal symmetric (TRS) Weyl semimetal film, the confinement gap gives the quasiparticles a mass, while TRS is preserved by having an even number of valleys with opposite masses. 
The film can be tuned through a topological phase transition by a gate electric field. In this work, we present a theoretical study of the quantum corrections to the conductivity of a topological semimetal thin film, which is governed by the complex interplay of the chiral band structure, mass term, and scalar and spin-orbit scattering. We study scalar and spin-orbit scattering mechanisms on the same footing, demonstrating that they have a strong qualitative and quantitative impact on the conductivity correction. We show that, due to the spin structure of the matrix Green's functions, terms linear in the extrinsic spin-orbit scattering are present in the Bloch and momentum relaxation times, whereas previous works had identified corrections starting from the second order.
In the limit of small quasiparticle mass, the terms linear in the impurity spin-orbit coupling lead to a potentially observable density dependence in the weak antilocalization correction.
Moreover, when the mass term is around 30 percent of the linear Dirac terms, the system is in the unitary symmetry class with zero quantum correction and switching the extrinsic spin-orbit scattering drives the system to the weak antilocalization.
We discuss the crossover between the weak localization and weak antilocalization regimes in terms of the singlet and triplet Cooperon channels, and analyze this transition as a function of the mass and spin-orbit scattering strength.  Experimental schemes to detect this transition are discussed.
\end{abstract}
\date{\today}
\maketitle
\end{CJK*}

\section{Introduction}

In 1929 Hermann Weyl theoretically predicted the existence of chiral massless fermions with a linear dispersion by identifying a specific solution of the Dirac equation. Over the past 80 years, the search for Weyl fermions has stimulated numerous studies in high energy, condensed matter, and mathematical physics. \cite{Balents11physics,%
Vishwanath15physics,Cianci15eurphysjc}
Recently, first-principles theoretical predictions followed by angle-resolved photo-emission spectroscopy (ARPES) experiments have confirmed the existence of chiral massless Weyl fermions \cite{Xiao10rmp} in topological Dirac semimetals,
\cite{Young12prl,Wang12prb,Wang13prb,Liu14science,Liu14natmater,Neupane14natcommun,%
Borisenko14prl,Yi14scirep,Xiong15science,Pixley15prl,Kushwaha15aplmater,Xiao15scirep,%
Li16natcommun,Burkov16prl,Xiong16epl,Zhao16scirep} as well as in type-I \cite{Wan11prb,Yang11prb,%
Singh12prb,Bulmash14prb,Xu15science,Weng15prx,Huang15natcommun,%
Lv15prx,Zhang16natcommun,Wu16apl} and type-II
\cite{Soluyanov15nature,Wang16natcommun,Deng16natphys,Wang16prb,Borisenko15arxiv}
Weyl semimetals (WSM).

\begin{figure}[t!]
\includegraphics[width=0.5\columnwidth]{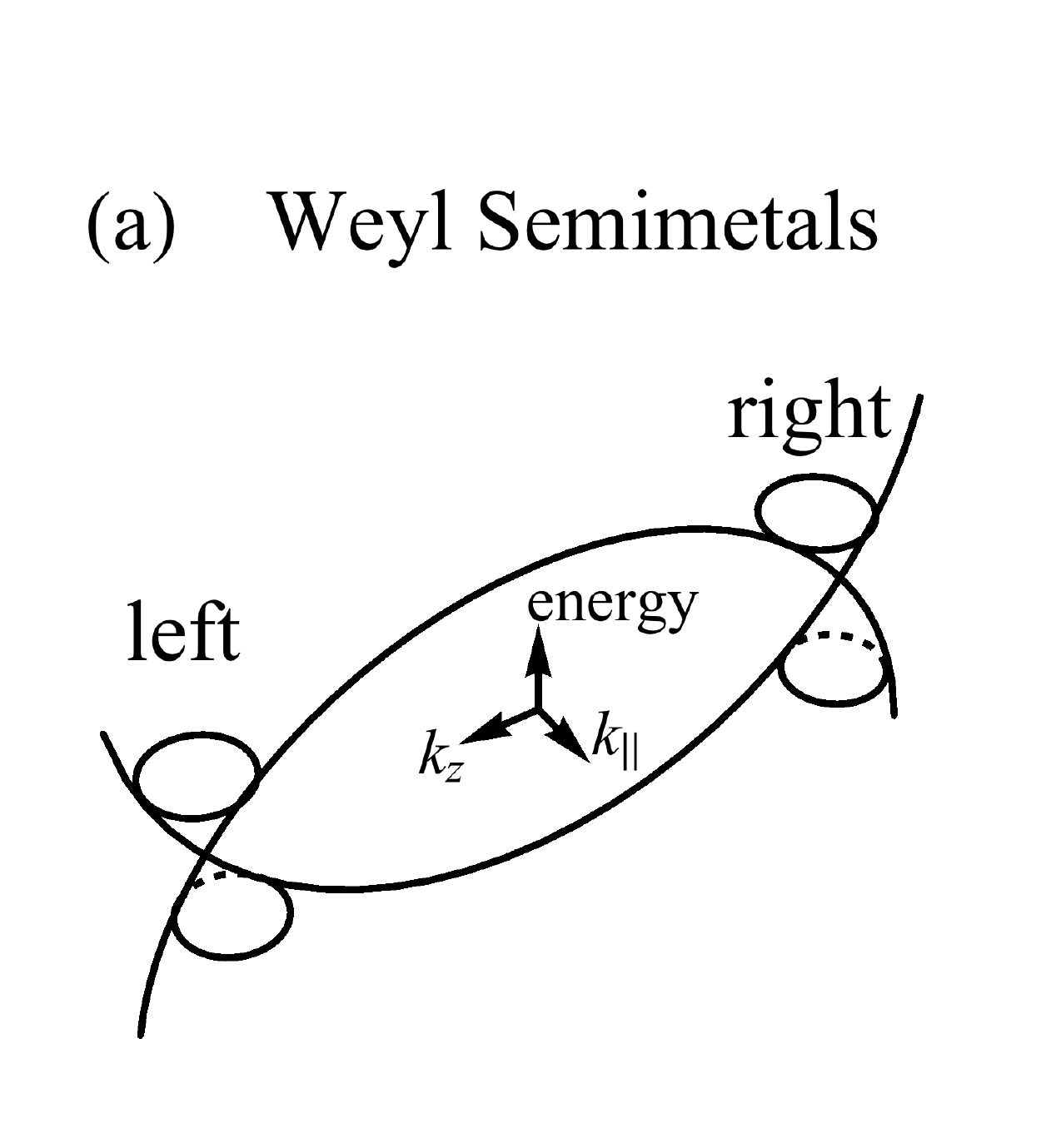}
\hspace{0.5cm}
\includegraphics[width=0.39\columnwidth]{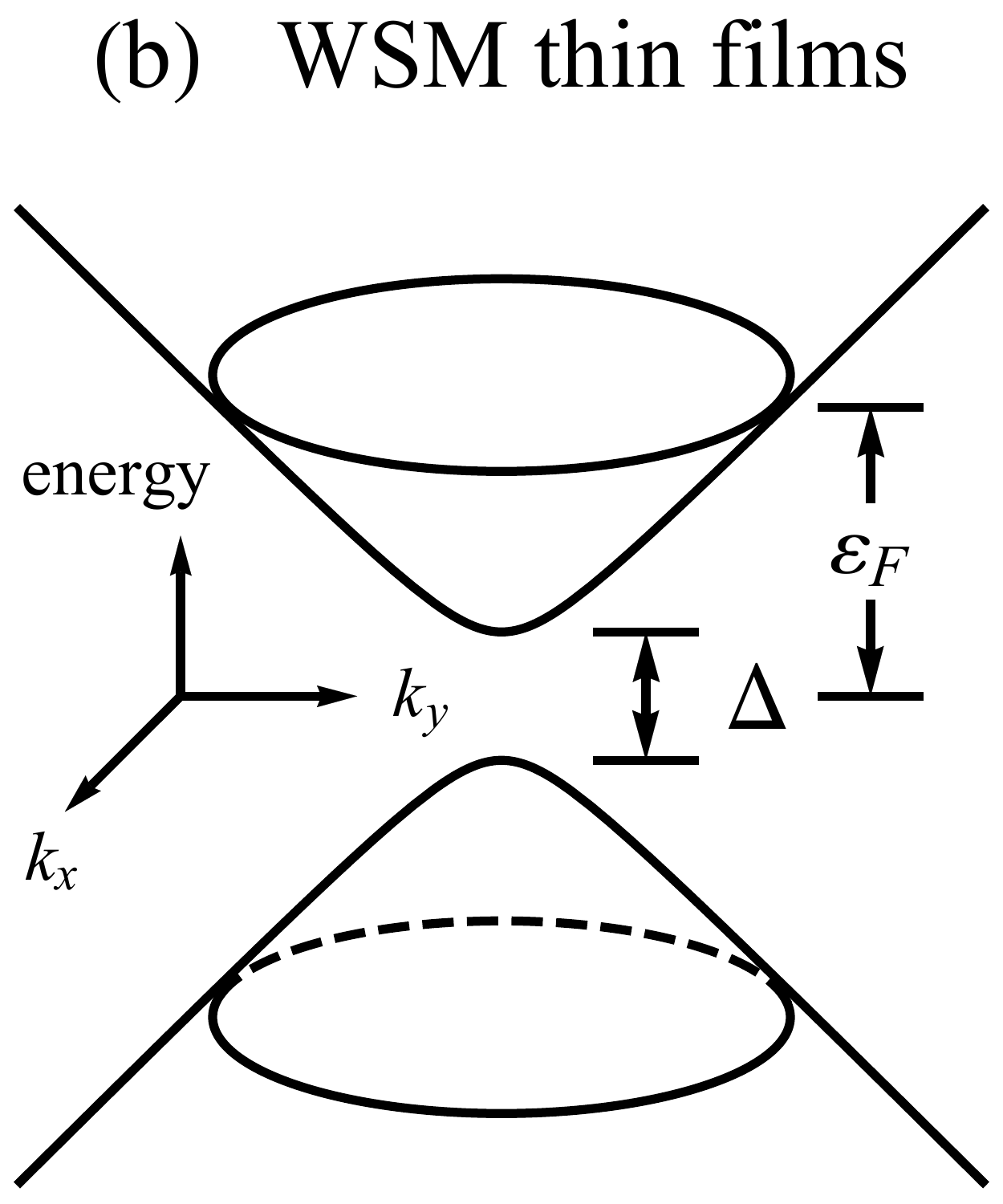}
\caption{\textbf{(a)} A minimal sketch of the energy dispersion of a Weyl semimetal. We have defined $k_{\parallel}^{2} = k_{x}^{2} + k_{y}^{2}$ while $(k_{x}, k_{y}, k_{z})$ represents the 3D wavevector.
$k_{z}$ points along the preferred direction.
The conductance and valence bands cross linearly at the Weyl nodes, i.e. the left and right cones, and the nodes appear in pairs with opposite chirality number. A Dirac node appears when two oppositely-chiral Weyl nodes merge together. \textbf{(b)} A schematic picture of the band structure in WSM thin films, where $\Delta$ is the band gap and $\varepsilon_{\text{F}}$ is the Fermi energy.}
\label{Weyl_basic}
\end{figure}

Weyl fermion semimetals are three-dimensional topological states of matter, \cite{Potter14natcommun,%
Burkov16natmater,Pixley16prx,Yang16spin}
in which the conduction and valence bands touch linearly at an even number of nodes [Fig.~\ref{Weyl_basic}(a)], which appear in pairs with opposite chirality. This band structure is referred to as topological \cite{Koenig07science,Culcer09prb,%
Moore10nature,Hasan10rmp,Qi10rmp,%
Culcer11prb,Li13cpb,Zhang14natcommun,Peng16natcommun,Liu16physicae,Liu16arxiv} because it is equivalent to a two-dimensional band insulator with the gap controlled by the momentum ($k_z$) along the direction connecting two nodes. As $k_z$ changes from outside to between the paired Weyl nodes, the system undergoes a topological phase transition from a trivial phase to a quantum anomalous Hall phase. \cite{Culcer10prb} As a result, there are $k_z$-dependent topological edge states, \cite{Oh13science} forming the Fermi arcs that connect the paired Weyl fermions. \cite{Xu15science2,Lau17arxiv} In an ultrathin film of topological Weyl semimetal (WSM), $k_z$ is quantized giving rise to a mass (or equivalently a gap $\Delta$), then the system [Fig.~\ref{Weyl_basic}(b)] exhibits precisely the same 2D massive Dirac fermion states \cite{Lu10prb,Abanin11prl} as those associated with the quantum anomalous Hall effect. \cite{Nagaosa10rmp,Burkov14prl} In a similar manner, a Dirac semimetal thin film may host the quantum spin Hall effect, since it can be regarded as consisting of two Weyl semimetals that are time-reversed pairs. \cite{Wang13prb} In contrast to previously studied systems, \cite{Yu10science,%
Chang13science,Yang09arxiv} the quantum spin and anomalous Hall effects may be observed in a single-compound device, a fact that has stimulated a considerable amount of recent interest in quantum transport in thin films of topological semimetals.

Topological semimetals can demonstrate a variety of transport phenomena.
\cite{Liu11prb,Burkov14prl2,Huang15prx,Lu15prb,Burkov15prb,%
Goswami15prb,Zhang16newjphys,Li17frontphys,Louvet17prb,Niemann16arxiv}
The ability to observe the quantum spin and anomalous Hall effects requires the Fermi energy to lie inside the gap opened by the quasiparticle mass. When this happens then a set of 1D edge states is well defined inside the gap, whose ballistic transport is responsible for the quantized conductance observed experimentally. When the Fermi energy is not in the mass gap, quantum transport at low temperatures is dominated by weak localization (WL) or antilocalization (WAL) effects.
\cite{Yang13prb,Lu16cpb,Kammermeier16prb} These corrections to the conductivity are noticeable when the quasiparticle mean free path is much shorter than the phase coherence length \cite{Evers08rmp} and arise as a result of the quantum interference between closed, time-reversed loops that circle regions in which one or more impurities are present. \cite{Lee85rmp} Since the interference effects leading to WL/WAL disappear in weak external magnetic fields these corrections can typically be identified straightforwardly in an experiment, and are frequently used to characterize samples, in particular transport in novel materials. They provide valuable information about the system, \cite{Koga02prl,Kim13natcommun} such as symmetries of the system, the phase coherence length, \cite{Muhlbauer14prl,Lu14prl} and the mass of the Dirac fermions.
\cite{Lu11prl}

Thin films of topological semimetals provide a new platform to understand weak localization and antilocalization behavior in generic 2D Dirac fermion systems in which the mass may be taken as a parameter. \cite{Imura09prb,%
He11prl,Tkachov11prb,Lu11prl,Yoshimura16prb} In the seminal work by Hikami, Larkin, and Nagaoka for conventional electrons (with a parabolic dispersion $\varepsilon_p = p^2/2m$), weak localization and antilocalization effects are classified according to the orthogonal, symplectic, and unitary symmetry classes, corresponding to scalar, spin-orbit, and magnetic impurities respectively. \cite{Hikami80progtheorphys} In the context of the 2D massive Dirac fermions, Shan, Lu, and Shen have considered all these impurity classes, \cite{Shan12prb} where weak spin-orbit scattering, in analogy with the treatment in Ref.~[\onlinecite{Hikami80progtheorphys}], is included by retaining only the second-order terms in the strength of the impurity spin-orbit coupling.

In this article, we formulate a complete theory describing weak localization and antilocalization of 2D massive Dirac fermions in topological semimetal thin films. We focus on a Weyl semimetal thin film as a prototype system. We stress that in these materials, in which the band structure spin-orbit interactions are exceedingly strong, spin-orbit coupling in the impurity scattering potentials is also expected to be sizable. It is, therefore, necessary to treat scalar and spin-dependent scattering on the same footing, which requires one to retain the matrix structure of all the Green's functions and impurity potentials. We have developed a transparent theory that accounts fully for the matrix structure of the Dirac/Weyl system within the framework of the Keldysh Green's function formalism. The resulting weak localization/antilocalization behavior is consistent with the universality classes in the massless and massive limits respectively and with the symmetries of the system under chirality reversal.

Thanks to the matrix structure inherent in our formalism, a major theoretical advance of this study compared to previous works \cite{Shan12prb} is our revelation of the existence of a linear term in the strength of the extrinsic spin-orbit scattering potential, which has a strong angular dependence. This term appears in the Bloch lifetime of the quasiparticles, in the transport relaxation time, the spin relaxation time, and the Cooperon, and gives rise to a non-trivial density dependence of the quantum correction to the conductivity, which may be observable when the quasiparticle mass is very small. Taking into account terms up to second-order in the extrinsic spin-orbit scattering we analyze quantitatively the carrier density dependence of the electrical conductivity and provide a fitting formula that may be used to extract the strength of the impurity spin-orbit interactions, as outlined in Sec.~\ref{carrierdensitydep}. Furthermore, the first-order spin-orbit scattering contribution suppresses the weak localization channel in the massive-limit much more efficiently than the second-order term studied previously, and therefore has a strong qualitative and quantitative effect on the phase diagram of the weak localization to weak antilocalization transition. With this insight, we determine the full phase diagram as a function of the strengths of the extrinsic spin-orbit coupling and of the mass term in the quasiparticle dispersion, which may be regarded as the central result of this work.

This paper is organized as follows. In Sec.~\ref{ModelFormalism}, we describe the generic model and theoretical formalism utilized throughout this article. In Sec.~\ref{Results}, we briefly display the conductivity dependence as a function of the external magnetic field, the carrier density, and the sign of the $\sigma_{y}$ term in the band Hamiltonian.
In Sec.~\ref{Discussions}, we discuss the physical implications of our results and their potential experimental applications. In Sec.~\ref{Conclusions}, we summarize our conclusions briefly and discuss possible future research directions.

\section{Model and Formalism}\label{ModelFormalism}

In this section, we will illustrate our model and develop our formalism by Keldysh Green's functions.
Then we will present expressions for the Drude conductivity and the quantum interference conductivity,
along with a solution to the Bethe-Salpeter equation in a matrix form.

\subsection{Band Hamiltonian}

In WSM thin films, the effective band Hamiltonian in the vicinity of a Weyl node takes the form:
\cite{Lu16frontphys,Culcer05prb}
\begin{equation}\label{WeylHamEq}
H_{0\bm{k}} = A (\sigma_{x} k_{x} + \sigma_{y} k_{y}) + M \sigma_{z},
\end{equation}
where $\sigma_{x,y,z}$ are Pauli matrices, $A = \hbar v$ is a material-specific constant, $v$ is the effective velocity, $\bm{k} = (k_{x}, k_{y})$ is the in-plane wavevector measured from the Weyl node, and $M$ is the effective mass due to the quantum confinement, which may also be viewed as an effective Zeeman energy. $H_{0\bm{k}}$ describes two chiral bands whose eigenvalues are $\displaystyle \varepsilon^{\pm}_{\bm{k}} = \pm \sqrt{A^{2} k^{2} + M^{2}} \equiv \pm \varepsilon_{k}$, where $+$ and $-$ indicate the conduction and valence bands, respectively. This is seen in Fig.~\ref{Weyl_basic}(b), in which the gap is $ \Delta = 2 M$. In this work, we assume that the Fermi level is located in the conduction band and $ \varepsilon \sim \varepsilon_{\text{F}} = \sqrt{A^{2} k_{\text{F}}^{2} + M^{2}}$ at low temperatures, where $k_{\text{F}} $ is the Fermi wavevector. Since transport properties only rely on the Fermi level, we introduce $a = A k_{\text{F}} / \varepsilon_{\text{F}}$ and $b = M / \varepsilon_{\text{F}}$ where $a^{2} + b^{2} = 1$.

\subsection{Impurity potential}\label{sec::ImpurityPotential}

For short-range impurities, the matrix elements of a single impurity potential in the reciprocal space, including spin-orbit-coupling, are given by
\begin{equation}\label{SOIP}
U_{\bm{k}\bm{k}^{\prime}} = \mathcal{U}_{\bm{k}\bm{k}^{\prime}} \left( \openone + i \lambda \sigma_{z} \sin \gamma \right),
\quad \gamma =  \theta^{\prime} - \theta,
\end{equation}
where $\mathcal{U}_{\bm{k} \bm{k}^{\prime}} \equiv \mathcal{U}$ is taken to represent the reciprocal-space matrix element of a short-range impurity potential, $\theta (\theta^{\prime})$ is the polar angle of the vector $\bm{k} (\bm{k}^{\prime})$, $\openone$ represents the $2\times2$ identity matrix and $i$ is the imaginary unit.
We assume that $\lambda < 1$ allowing us to do perturbation theory in this parameter in what follows. In this work we shall use a simplified notation for the spin-orbit contribution to the impurity potential as introduced in Eq.~\ref{SOIP}, however we note that $\lambda \sigma_{z} \sin \gamma$ represents a term that is frequently written as $\lambda_0 {\bm \sigma} \cdot {\bm k} \times {\bm k}'$, where $\lambda_0$ is a material-specific constant and $|{\bm k}| = |{\bm k}'| = k_F$. Hence it is important to emphasize the fact that $\lambda \propto k_{\text{F}}^{2} = 4 \pi n_{e}$ is a linear function of the electron density $n_e$, which experimentally can be tuned by changing the gate voltage or the temperature. \cite{Shekhar15natphys}
The total impurity potential in the real space is 
\begin{eqnarray}
V(\bm{r}) = \sum_{I} U (\bm{r} - \bm{R}_{I}),
\end{eqnarray}
where $\sum_{I}$ is a summation over all impurities and $\bm{R}_{I}$ is the impurity coordinate.
Following an average over random impurity configurations,
the total impurity potential in the reciprocal space $V_{\bm{k}\bm{k}^{\prime}}$ becomes $\overline{V^{\alpha\beta}_{\bm{k}\bm{k}^{\prime}}} = 0$ and $\overline{V_{\bm{k}\bm{k}^{\prime}}^{\alpha\beta} V_{\bm{k}_{1}\bm{k}_{1}^{\prime}}^{\eta\zeta}} = n_{i} U^{\alpha\beta}_{\bm{k}\bm{k}^{\prime}} U_{\bm{k}_{1}\bm{k}_{1}^{\prime}}^{\eta\zeta}$
where $ \alpha,$ $\beta,$ $\eta,$ and $\zeta$ are spin indices, and  $n_{i}$ representing the impurity concentration.
For strong screening \cite{Ando06jphyssocjpn,Aleiner06prl,Adam12prb,DasSarma15prb} we may assume $\mathcal{U} = Z / N_{\text{F}},$ where $N_{\text{F}} = \varepsilon_{\text{F}} /(2 \pi A^{2})$ is the density of states at $\varepsilon_{\text{F}}$ and $Z \equiv 1$ is the atomic number of the impurities.

\begin{figure}[b]
\begin{center}
\includegraphics[width=0.35\columnwidth]{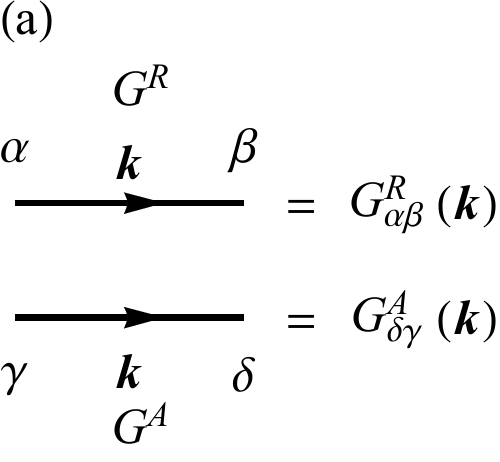} \quad
\includegraphics[width=0.45\columnwidth]{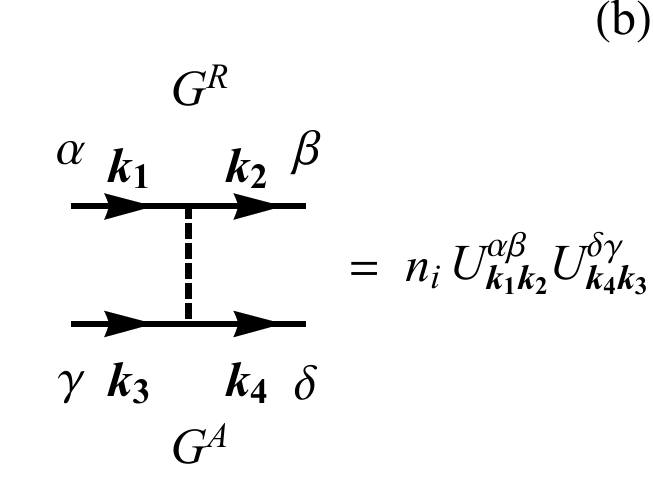} \\
\includegraphics[width=0.3\columnwidth]{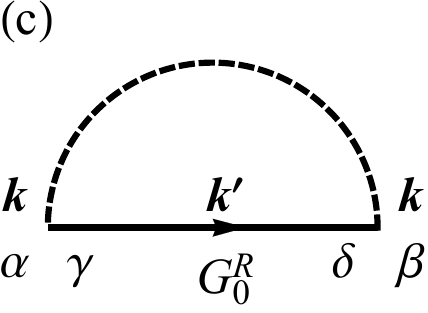} \quad \quad
\includegraphics[width=0.5\columnwidth]{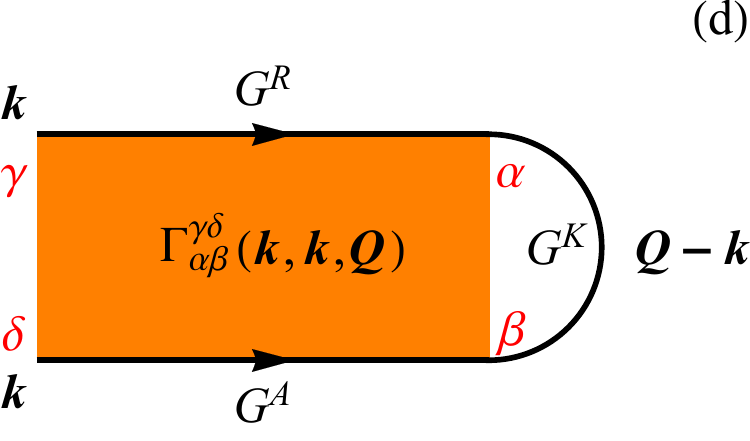} \\
\includegraphics[width=0.9\columnwidth]{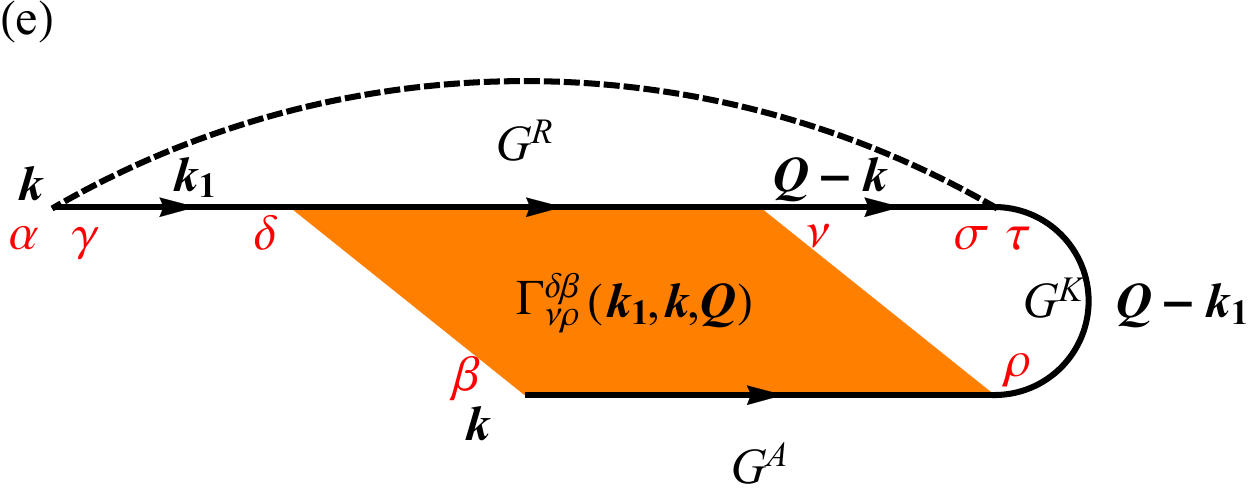} \\
\includegraphics[width=\columnwidth]{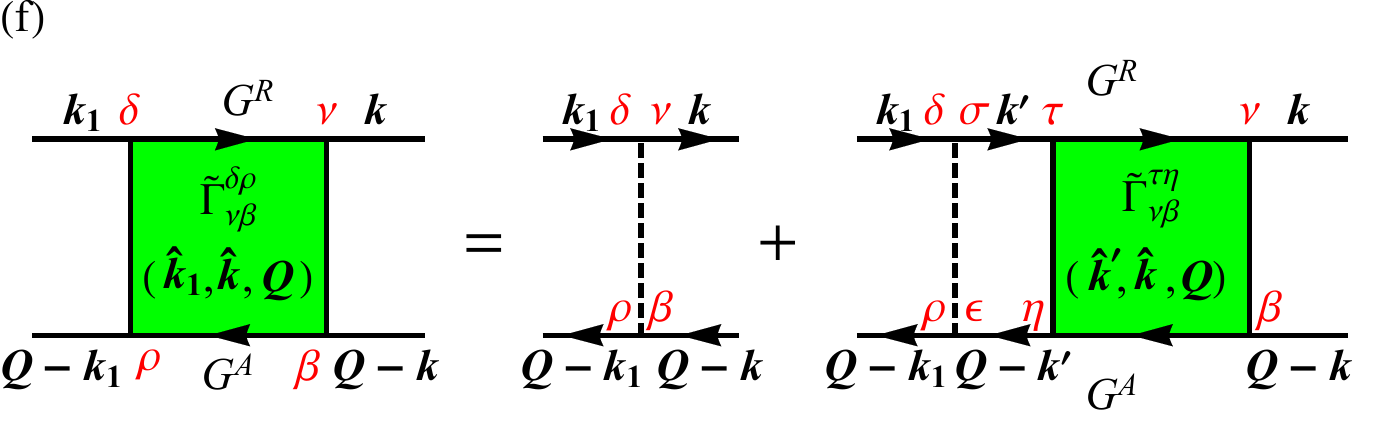}
\caption{(Color online) The diagrams for the weak (anti-)localization conductivity of Dirac fermions.
(a) Define Green's function as arrowed solid lines in which Greek letters are spin indices.
(b) Definition of dashed lines: impurities lines expressed in both retarded and advanced cases.
(c) The retarded self-energy in the first-order Born approximation,
where $G^{\text{R}}_{0}$ is the bare retarded Green's function.
(d) and (e) are Keldysh self-energies $\big(\Sigma^{\text{K},\text{b}}_{\bm{k},\gamma\delta}$ and
$\Sigma^{\text{K},\text{R}}_{\bm{k},\alpha\beta} \big)$
of maximally crossed diagrams in the bare and the retarded dressed cases, respectively,
where $\Gamma$ is the Cooperon structure factor.
(f) The Bethe-Salpeter equation for the twisted Cooperon structure factor $\tilde{\Gamma}.$}
\label{All_Diagrams}
\end{center}
\end{figure}

\subsection{Summary of the Keldysh formalism}

In the following, we will follow a version of the well-established formalism due to Keldysh\cite{Keldysh65sovphysjetp} to calculate the electrical conductivity. The central quantity in the Keldysh formalism is the Keldysh Green's function $G^{\text{K}}$ whose physical content is analogous to that of the density matrix $\rho$ in the quantum Liouville equation. In the Keldysh representation, the Keldysh Green's function $G^{\text{K}}$, together with the retarded and advanced Green's functions $G^{\text{R}}$ and $G^{\text{A}}$, form a $2\times2$ matrix:
\begin{equation}
\check{G} = \left(
\arraycolsep 1ex
\begin{array}{cc}
0 & G^{\text{A}} \\[2ex]
G^{\text{R}} & G^{\text{K}}
\end{array}
\right).
\end{equation}
The Green's function matrix $\check{G}$ is related to the \textit{contour-ordered} Green's function $\hat{G}$ through the equation 
\begin{eqnarray}
 \hat{G} = R \, \check{G} \, R^{-1},
\end{eqnarray}
with $R = (1+i\sigma_{y})/\sqrt{2}$, where the contour takes into account time evolution in both directions. In real space, the time evolution of $\hat{G}$ is given by
\begin{eqnarray}
 \left( i \hbar \, \partial_{t_{1}} - H_{\bm{r}_{1} t_{1}} \right) \hat{G} (1,2) &=& \sigma_{z} \delta (1-2) , \\[1ex]
 \hat{G} (1,2) \left( - i \hbar \, \overleftarrow{\partial}_{t_{2}} - \overleftarrow{H}_{\bm{r}_{2} t_{2}}\right)
&=& \sigma_{z} \delta (1 - 2),
\end{eqnarray}
where $\partial_{t_{1}} \equiv \partial / \partial t_{1}$ and the overhead left-arrow means that the derivative acts on the left part. Here we use abbreviations $1\equiv (t_{1}, \bm{r}_{1})$ and $\delta (1-2) = \delta (\bm{r}_{1} - \bm{r}_{2}) \delta (t_{1} - t_{2})$, while $H = H_{0} + V $ is the total Hamiltonian, with $H_0$ the band Hamiltonian and $V$ the total impurity potential as introduced above.

The Dyson equation for $\hat{G}$ may be written as
\begin{equation}
\hat{G} (1,2) = \hat{g} (1,2) + \int \text{d} 3 \int \text{d} 4  \, \hat{g} (1,3) \hat{\Sigma} (3,4) \hat{G} (4,2).
\end{equation}
where $\text{d} 3 = \text{d} \bm{r}_{3} \, \text{d} t_{3}$ etc. The self-energy matrix $\hat{\Sigma}$ is
\begin{equation}
\hat{\Sigma} = R \left(
\begin{array}{cc}
\Sigma^{\text{K}} & \Sigma^{\text{R}} \\
\Sigma^{\text{A}} & 0
\end{array}\right) R^{-1},
\end{equation}
where $\Sigma^{\text{K}},$ $\Sigma^{\text{R}},$ and $\Sigma^{\text{A}}$ are the Keldysh, retarded, and advanced self-energies.
After the average over impurity configurations, the retarded component of the self-energy in the Born approximation is shown in Fig.~\ref{All_Diagrams}(c),
while the Keldysh components of the self-energy leading to quantum-interference are in Figs.~\ref{All_Diagrams}(d) and \ref{All_Diagrams}(e).\cite{Vasko05book}
In Fig.~\ref{All_Diagrams}, the spin indices are kept, not the indices in the Green's function matrix. The diagrams appearing in Fig.~\ref{All_Diagrams} will be important for the upcoming calculations of the Drude conductivity and the quantum-interference correction to the conductivity.

Based on the above, omitting a series of cumbersome technical details, in an external electric field $\bm{E}$ that is constant and uniform, the quantum kinetic equation for $G^{\text{K}}$ may be written in the form \cite{Rammer86rmp}
\begin{equation}\label{Kineticeq}
\partial_{t} G^{\text{K}}_{\varepsilon {\bm k}} + ( i / \hbar ) \big[ H_{0{\bm k}} , G^{\text{K}}_{\varepsilon {\bm k}}\big]
+ \mathcal{J}_{\varepsilon {\bm k}} = (e/ \hbar){\bm E} \cdot \partial_{\bm{k}} G^{\text{K}}_{\varepsilon {\bm k}},
\end{equation}
where $\partial_{\bm{k}} \equiv \partial / \partial {\bm k}$, $e$ is the elementary charge (thus the electron charge is $-e$), $\varepsilon$ is an intermediate energy variable that is integrated over (representing non-locality in time), and the scattering term $\mathcal{J}_{\varepsilon{\bm k}}$ is given by
\begin{equation}\label{J_epsilon_k}
\mathcal{J}_{\varepsilon{\bm k}} \!= \! ( i / \hbar ) \big(\Sigma^{\text{R}}_{\varepsilon{\bm k}}
G^{\text{K}}_{\varepsilon{\bm k}} - G^{\text{K}}_{\varepsilon{\bm k}} \Sigma^{\text{A}}_{\varepsilon{\bm k}}
+ \Sigma^{\text{K}}_{\varepsilon{\bm k}} G^{\text{A}}_{\varepsilon{\bm k}}
- G^{\text{R}}_{\varepsilon{\bm k}} \Sigma^{\text{K}}_{\varepsilon{\bm k}} \big).
\end{equation}
We adopt the form $\displaystyle G^{\text{K}}_{\varepsilon {\bm k}} = \chi_{\bm k} (G^{\text{R}}_{\varepsilon {\bm k}}$ $- G^{\text{A}}_{{\varepsilon}{\bm k}} )$ with $\chi_{\bm k}$ a scalar, following the reasoning of Ref.~[\onlinecite{Schwab11epl}]. Note that the Wigner transformation is applied on $\hat{G}(1,2)$ in order to find the single-particle Green's function $\hat{G}_{\varepsilon\bm{k}}$.

In the following, we will introduce the disorder averaged Green's functions in Sec.~\ref{DAGF}, the Drude conductivity in Sec.~\ref{DrudeCon}, and quantum-interference conductivity in Sec.~\ref{QuantumInterCon}.

\subsection{Self-Consistent Retarded and Advanced Green's functions}\label{DAGF}

The arrowed lines in Fig.~\ref{All_Diagrams} are Green's functions that are defined in Fig.~\ref{All_Diagrams}(a),
and the dashed lines in Fig.~\ref{All_Diagrams} corresponds to the impurity scattering
whose definition in the retarded and the advanced cases are displayed in
Fig~\ref{All_Diagrams}(b). The bare Green's function is
\begin{equation}
G_{0\bm{k}} = \frac{\varepsilon \openone + H_{0\bm{k}}}{\varepsilon^{2} - \varepsilon_{k}^{2}}
\xrightarrow{\varepsilon \to \varepsilon_{k}}
\frac{\openone + H_{0\bm{k}} / \varepsilon_{k} }{2(\varepsilon - \varepsilon_{k})}
= \frac{g_{\bm{k}}}{\varepsilon - \varepsilon_{k}},
\end{equation}
where $g_{\bm{k}} = [ \openone + H_{0\bm{k}} / \varepsilon_{k} ]/2.$
Under the first-order Born approximation, the retarded self-energy
$\Sigma^{\text{R}}_{\text{Bn}} $ is expressed in Fig.~\ref{All_Diagrams}(c) and the disorder-averaged
(retarded and advanced) Green's functions are
\begin{equation}
\hspace{-0.18cm} G^{\text{R}}_{\bm{k}} = \big[ \openone - G^{\text{R}}_{0\bm{k}} \Sigma^{\text{R}}_{\text{Bn}} \big]^{-1}G^{\text{R}}_{0\bm{k}}
= \frac{g_{\bm{k}}}{\varepsilon - \varepsilon_{k} + \frac{i \hbar}{2\tau} } = \big[ G^{\text{A}}_{\bm{k}} \big]^{*},
\end{equation}
where $G^{\text{R}}_{0\bm{k}} = g_{\bm{k}} / (\varepsilon - \varepsilon_{\bm{k}} + i\, 0^{+})$, $^*$ denotes the Hermitian conjugate, and $G^{\text{A}}_{\bm{k}}$ is the advanced Green's function. The elastic scattering time $\tau$ is
\begin{equation}\label{tau}
\tau = \tau_{0} / \big[ \big( 1+ \lambda^{2}/2 \big) \big( 1 + b^{2} \big)
+ \lambda a^{2} \big],
\end{equation}
where $1/\tau_{0} = \pi n_{i} |\mathcal{U}|^{2} N_{\text{F}} /\hbar.$
In Eq.~(\ref{tau}), the linear $\lambda$ term comes from the chirality interplay
between the band Hamiltonian and the spin-orbit impurities and only appears when we maintain the matrix structure.
Compared with the topological insulator case, \cite{Adroguer15prb}
the opposite chirality number in band Hamiltonian gives the opposite sign of the linear term.

We have retained the matrix structure of all Green's functions and note that: (i) a linear-in-$\lambda$ term appears in the elastic scattering time $\tau$ and transport time $\tau_{\text{tr}}$, which is missed if the matrix structure is neglected, and has not been discussed previously in Dirac/Weyl semimetals;
(ii) a non-trivial coupling between the singlet and the triplet channels in the Cooperon emerges below when first-order impurity spin-orbit coupling terms are taken into account in backscattering processes;
(iii) in order to obtain the correct result in (i) and (ii), it is necessary to consider explicitly the angular dependence of the spin-orbit coupling terms in the impurity potential, in contrast to the conventional Hikami-Larkin-Nagaoka approach in which the square of these terms is averaged over the Fermi surface.

\subsection{Drude conductivity}\label{DrudeCon}

With the self-energy in the Born approximation as shown in Fig.~\ref{All_Diagrams}(c), the Born-approximation scattering term becomes
$\mathcal{J}_{\text{Bn}} (\chi_{\bm{k}}) = - 2 \, i \pi \, g_{\bm{k}} \, \chi_{\bm{k}} / \tau_{\text{tr}}$, where the transport scattering time (momentum relaxation time) is
\begin{equation}\label{trsptime}
\tau_{\text{tr}} = 2 \tau_{0} / \big[1 + 3 b^{2} + 2 a^{2} \lambda +
\big( 5 + 3 b^{2}\big) \lambda^{2} /4 \big].
\end{equation}
The elastic scattering length is $\ell_{e} = \sqrt{D \tau}$ where $D = v_{\text{F}}^{2} \tau_{\text{tr}} / 2$ is the diffusion constant and $v_{\text{F}} = a v$ is the Fermi velocity.
We assume that $\ell_{e}$ is much shorter than the phase coherence length $\ell_{\phi}$ as required in diffusive quantum transport.
To the leading order of $\tau_{\text{tr}}^{-1}$ the solution of Eq.~(\ref{Kineticeq}) is $\chi_{E\bm{k}} = [2 e \bm{E} \cdot \hat{\bm{k}} \, \tau_{\text{tr}} / \hbar] \delta (k - k_{\text{F}})$
where $\hat{\bm{k}}$ is the unit vector along $\bm{k}$, and the Drude conductivity takes the form
\begin{equation}
\sigma^{\text{Dr}}_{xx} = (e^{2}/h) (v a k_{\text{F}} \tau_{\text{tr}} / 2),
\end{equation}
where $h$ is the Planck constant.

\subsection{Quantum interference correction to the conductivity}\label{QuantumInterCon}

The quantum-interference between two time-reversed closed trajectories will generate three different contributions
to the conductivity, whose Keldysh self-energies are (bare case) $\Sigma^{\text{K}}_{\text{b}},$
(retarded dressed case) $\Sigma^{\text{K}}_{\text{R}},$
and (advanced dressed case) $\Sigma^{\text{K}}_{\text{A}}.$
The Keldysh self-energies diagrams of $\Sigma^{\text{K}}_{\text{b}}$ and $\Sigma^{\text{K}}_{\text{R}}$
are shown in Figs.~\ref{All_Diagrams}(d) and \ref{All_Diagrams}(e), respectively.
In these diagrams, the maximally crossed diagrams
(Cooperon structure factor) $\Gamma$ is proportional to $1/|\bm{Q}|^{2},$
where $\bm{Q}$ is the sum of the incoming and the outgoing wavevector.
The divergence of $\Gamma_{\bm{Q}}$ at $|\bm{Q}| \to 0$ indicates the primary contribution into
the quantum-interference conductivity comes from the backscattering.
The quantum-interference self-energies ($\Sigma^{\text{K}}_{\text{b}},$ $\Sigma^{\text{K}}_{\text{R}}$
and $\Sigma^{\text{K}}_{\text{A}}$) will generate the quantum-interference scattering term
$\mathcal{J}_{\text{qi}} (\chi_{E\bm{k}})$ that is balanced by $\mathcal{J}_{\text{Bn}} (\chi_{\text{qi},\bm{k}})$ as follows:
\begin{equation}
\mathcal{J}_{\text{Bn}} (\chi_{\text{qi},\bm{k}}) = - \mathcal{J}_{\text{qi}} (\chi_{E\bm{k}}),
\end{equation}
where the right-hand side plays the role of a driving term. The term $\chi_{\text{qi},\bm{k}}$ found from this equation leads to the quantum-interference conductivity correction
\begin{equation}\label{sigmaWAL_Ini}
\sigma^{\text{qi}}_{xx} = - ( e^{2} v_{\text{F}}^{2} \tau^{3} N_{\text{F}} \eta_{v}^{2} / \hbar^{2} )
\int_{\bm{Q}} \big[ C^{\text{b}}_{\bm{Q}} + 2 C^{\text{R}}_{\bm{Q}} \big],
\end{equation}
where $\int_{\bm{Q}} \equiv \int \text{d} \bm{Q} / (2 \pi)^{2}$, and $\eta_{v} = \tau_{\text{tr}} / \tau$ is the
current vertex renormalization factor.
$ C^{\text{b}}_{\bm{Q}}$ and $ C^{\text{R}}_{\bm{Q}}$
are the bare Cooperon and the retarded-dressed Cooperon, respectively,
where $ C_{\bm{Q}} = \text{Re} \big[ \Sigma^{\text{K}}_{\bm{k},
\gamma\alpha}  g_{\bm{k}}^{\alpha \gamma}  \big]$ in general,
and $\alpha$ and $\gamma$ are spin indices.
The Einstein summation rule over repeated indices is used throughout this work.
Note that the advanced dressed Cooperon contributes the exactly same amount as its retarded dressed counterpart.

In order to solve for $\sigma^{\text{qi}}_{xx}$ explicitly, the Cooperon structure factor is twisted,
and the resulting structure factor satisfies
$\tilde{\Gamma}^{\delta\rho}_{\nu\beta} (\bm{k}_{1},\bm{k},\bm{Q})
= \Gamma^{\delta \beta}_{\nu \rho} (\bm{k}_{1},\bm{Q}-\bm{k},\bm{Q}),$
since the time-reversal symmetry is preserved and here we reverse the $G^{A}$ lines.
$\delta,$ $\nu,$ $\beta,$ and $\rho$ are spin indices.
From Fig.~\ref{All_Diagrams}(f), the twisted Cooperon structure factor $\tilde{\Gamma}$
is ladder-like and satisfies the Bethe-Salpeter equation~:
\begin{equation}\label{BetheSalpeter}
\tilde{\Gamma}_{mn} = b_{-m} \delta_{m,-n} + b_{-m} N_{-m + l} \, \tilde{\Gamma}_{ln},
\end{equation}
where $\tilde{\Gamma} (\bm{k}_{1} , \bm{k}, \bm{Q})$ is expanded as
$\tilde{\Gamma}_{mn} \text{e}^{im \theta_{1} + i n \theta}$
with $\theta_{1} =\theta_{\bm{k}_{1}},$ $ N_{n} = \int_{\bm{k}} e^{i n \theta_{\bm{k}}} \,
G^{\text{R}}_{\bm{k}} \otimes G^{\text{A}}_{\bm{Q} - \bm{k}},$ and
$b (\bm{k}_{1}, \bm{k},\bm{Q}) \approx \overline{U_{\bm{k}_{1}\bm{k}} \otimes U_{ - \bm{k}_{1}, - \bm{k}}}
= b_{n} \text{e}^{i n (\theta - \theta_{1})}$ is the bare vertex.
In Eq.~(\ref{BetheSalpeter}), we neglected the spin indices and
introduced $[A \otimes B]^{\alpha\beta}_{\delta\gamma} = A_{\alpha\delta} B_{\beta\gamma}$ for simplicity.
The leading solution in Eq.~(\ref{BetheSalpeter}) is $\tilde{\Gamma}_{00}$ and, up to $\mathcal{O} (\lambda^{2}),$
all angular-dependent Cooperon structure factor components are listed in Tab.~\ref{Gammanmexp}.
It is more convenient to transform Eq.~(\ref{BetheSalpeter}) into the singlet-triplet (ST) basis and
then to extract the singlet and triplet contributions separately. Also, note that it is crucial to include all off-diagonal terms in $[\tilde{\Gamma}_{00}/ b_{0}]^{-1},$ which are related
to the correlations between singlet and triplet channels.

\begin{table}[t!]
\caption{Up to the order of $\lambda^{2}$, all non-vanishing angular-dependent Cooperon
structure factor components $\Gamma_{mn}$ with their expressions.}
\begin{center}
\begin{tabular}{c|c||c|c}
$\mathcal{O} (\lambda)$ & Expr. & $\mathcal{O} (\lambda^{2})$ & Expr. \\
\hline
\multirow{2}{*}{$\Gamma_{-1,0}$} &
\multirow{2}{*}{$ \begin{array}{l} [ (\openone + b_{1} N_{0}) b_{1} N_{1} \\
\hspace{0.3cm} + b_{1} N_{2} b_{-1} N_{-1} ] \Gamma_{00} \end{array} $} &
$\Gamma_{-1,-1}$ & $b_{1} N_{1} \Gamma_{0,-1}$ \\[0.5ex]
\cline{3-4}
& & $\Gamma_{-1,1}$ & $b_{1} N_{1} \Gamma_{0,1}$ \\[0.5ex]
\hline
\multirow{2}{*}{$\Gamma_{1,0}$} &
\multirow{2}{*}{$ \begin{array}{l} [ (\openone + b_{-1} N_{0}) b_{-1} N_{-1} \\
\hspace{0.5cm} + b_{-1} N_{-2} b_{1} N_{1} ] \Gamma_{00} \end{array} $ } &
$\Gamma_{1,1}$ & $b_{-1} N_{-1} \Gamma_{0,1}$ \\[0.5ex]
\cline{3-4}
& & $\Gamma_{1,-1}$ & $b_{-1} N_{-1} \Gamma_{0,-1}$ \\[0.5ex]
\hline
\multirow{2}{*}{$\Gamma_{0,1}$} &
\multirow{2}{*}{$ \begin{array}{l} \Gamma_{00} [N_{-1} b_{1} (\openone + N_{0} b_{1}) \\
\hspace{0.7cm}+ N_{1} b_{-1} N_{-2} b_{1} ] \end{array} $} &
$\Gamma_{2,0}$ & $b_{-2} N_{-2} \Gamma_{00}$ \\[0.5ex]
\cline{3-4}
& & $\Gamma_{-2,0}$ & $b_{2} N_{2} \Gamma_{00}$ \\[0.5ex]
\hline
\multirow{2}{*}{$\Gamma_{0,-1}$} &
\multirow{2}{*}{$ \begin{array}{l} \Gamma_{00} [N_{1} b_{-1} (\openone + N_{0} b_{-1}) \\
\hspace{0.7cm} + N_{-1} b_{1} N_{2} b_{-1} ] \end{array} $} &
$\Gamma_{0,-2}$ & $\Gamma_{00} N_{2} b_{-2}$ \\[0.5ex]
\cline{3-4}
& & $\Gamma_{0,2}$ & $\Gamma_{00} N_{-2} b_{2} $ \\[0.5ex]
\end{tabular}
\end{center}
\label{Gammanmexp}
\end{table}%

\textit{Zero-field conductivity.}---The zero-field conductivity can
be obtained by integrating $Q \equiv |\bm{Q}|$ between $1/\ell_{e}$ and $1/\ell_{\phi}:$
\begin{equation}\label{zerofieldcondu}
\sigma^{\text{qi}}_{xx} (0) = \sum_{i=1,2,3} \frac{\alpha_{i} e^{2}}{\pi h} \ln
\frac{1/\ell_{i}^{2} + 1/\ell_{\phi}^{2}}{1/\ell_{i}^{2} + 1/\ell_{e}^{2}},
\end{equation}
where $i = 1, 2,$ and 3 are the singlet, the triplet-up, and the triplet-down channel indices, respectively,
$\alpha_{i} = \alpha_{i j} \lambda^{j} $ is the weight of channel $i,$
$1/\ell_{i}^{2} =\gamma_{i} / (v_{\text{F}} \tau)^{2}$
is the effective coherence length with $\gamma_{i} = \gamma_{i j} \lambda^{j}.$
The explicit expressions of $\alpha_{i,j}$ and $\gamma_{i,j}$ are listed in Tab.~\ref{alphagammaexp}.
In the massless limit $(b = 0),$
$\alpha_{1} = -1/2$ and $\gamma_{1} = 0$, matching the findings of Ref.~[\onlinecite{Adroguer15prb}].

\begin{table}[t!]
\caption{Explicit expressions of $\alpha_{i,j}$ and $\gamma_{i,j}$ components.
The definitions of $f_{i} (b)$ are
$ f_{1} (b) = [67 + 227 b^{2} + 143 b^{4} - 225 b^{6} - 679 b^{8} - 1127 b^{10} - 427 b^{12} - 27 b^{14}]/(1 + b^{2})^{4},$
$ f_{2} (b) = [11 - 252 b^{2} - 56 b^{3} - 58 b^{4} - 16 b^{5} + 428 b^{6} - 24 b^{7} - 289 b^{8}]/(1 + b^{2}),$
$ f_{3} (b) = 7 + 5 b + 5 b^{2} - b^{3},$
$ f_{4} (b) = [ 4 + 7 b + 11 b^{2} + 5 b^{3} + 5 b^{4} ]/(1 + b^{2}),$
$ f_{5} (b) = 41 + 154 b^{2} - 17 b^{4} - 52 b^{6} - 729 b^{8} - 358 b^{10} - 63 b^{12},$
and $f_{6} (b) = [43 - 152 b + 215 b^{2} - 688 b^{3} - 170 b^{4} - 64 b^{5} - 890 b^{6} + 944 b^{7} - 641 b^{8}
- 808 b^{9} + 163 b^{10}]/[(1 + b^{2})^{2} (1+b)^{2}].$}
\begin{center}
\begin{tabular}{c|c|c|c}
& $j=0$ & $j=1$ & $j=2$ \\
\hline
$\alpha_{1,j}$ & $\displaystyle -\frac{a^{4}}{2(1+3b^{2})}$ &
$\displaystyle -\frac{4 a^{2} b^{2} (3 + b^{2})}{(1+3b^{2})^{3}}$ &
$\displaystyle \frac{b^{2} f_{1} (b)}{(1+3b^{2})^{4}}$ \\
\hline
$\alpha_{2,j}$ & $\displaystyle \frac{4 b^{2} (1 + b^{2})}{(1+3b^{2})^{2}}$ &
$\displaystyle -\frac{2(1-b)b^{2} f_{3} (b)}{(1+3b^{2})^{3}}$ &
$\displaystyle \frac{b^{2} f_{2} (b)}{(1+3b^{2})^{4}}$ \\
\hline
$\alpha_{3,j}$ & $\displaystyle \frac{4 b^{2} (1 + b^{2})}{(1+3b^{2})^{2}}$ &
$\displaystyle -\frac{2(1+b)b^{2} f_{3} (-b)}{(1+3b^{2})^{3}}$ &
$\displaystyle \frac{b^{2} f_{2} (-b)}{(1+3b^{2})^{4}}$ \\
\hline
$\gamma_{1,j}$ & $\displaystyle \frac{2a^{2} b^{2}}{(1 + b^{2})^{2}}$ &
$\displaystyle -\frac{2 b^{2} (1-14 b^{2} - 3 b^{4})}{(1 + b^{2})^{3}}$ &
$\displaystyle - \frac{b^{2} f_{5} (b)}{2 a^{2} (1 + b^{2})^{6}}$ \\
\hline
$\gamma_{2,j}$ & $\displaystyle \frac{8(1 - b)^{2} b^{2}}{(1 + b)^{2} (1 + 3 b^{2})} $ &
$\displaystyle \frac{8(1-b) b^{3} f_{4} (b)}{(1 + b)^{2}(1+3b^{2})^{2}} $ &
$\displaystyle - \frac{b^{2} f_{6} (b)}{(1 + 3 b^{2})^{3}} $ \\
\hline
$\gamma_{3,j}$ & $\displaystyle \frac{8(1 + b)^{2} b^{2}}{(1 - b)^{2} (1 + 3 b^{2})} $ &
$\displaystyle - \frac{8(1+b) b^{3} f_{4} (-b)}{(1 - b)^{2}(1+3b^{2})^{2}} $ &
$\displaystyle - \frac{b^{2} f_{6} (-b)}{(1 + 3 b^{2})^{3}} $
\end{tabular}
\end{center}
\label{alphagammaexp}
\end{table}%

\textit{Magnetoconductivity formula.}---In an out-of-plane magnetic field $B$, $Q^{2}$ will be quantized as $Q_{n}^{2}
= (n + 1/2) / \ell_{B}^{2},$ where $n$ is the quantization number and $ l_{B} = \sqrt{\hbar / 4 e B}.$ Summing over $n$ from
$\ell_{e}^{2} / \ell_{B}^{2}$ to $\ell_{\phi}^{2} / \ell_{B}^{2}$ gives rise to the
field-dependent conductivity $\sigma^{\text{qi}}_{xx} (B).$ Since $\sigma^{\text{Dr}}_{xx}$ is nearly independent of $B$,
the magnetoconductivity $\Delta \sigma (B) = \sigma^{\text{qi}}_{xx}(B) - \sigma^{\text{qi}}_{xx} (0)$
is described by a general expression \cite{Hellerstedt16nanolett}
\begin{equation}\label{magnetoconduc}
\Delta \sigma (B) = \sum_{i =1,2,3} \frac{\alpha_{i} e^{2}}{\pi h}
\bigg( \Psi \bigg[ \frac{B_{\phi,i}}{ | B |} + \frac{1}{2} \bigg] - \ln \bigg[ \frac{B_{\phi,i}}{ | B | } \bigg] \bigg),
\end{equation}
where $B_{\phi,i} = (\hbar/4e) (1/\ell_{\phi}^{2} + 1/\ell_{i}^{2})$ is the effective
phase coherence field, and $\Psi$ is the digamma function.
When $x \ll 1,$ $\displaystyle y (x) = \Psi ( 1/ | x | + 1/2 ) - \ln \left( 1 / | x | \right) \propto x^{2},$
and when $x \gg 10,$ $y (x) \propto \sqrt{x}$. \cite{Lu16frontphys}

\section{Results}\label{Results}

In the following, the results for the magnetoconductivity, zero-field conductivity
and sign reversal of the $\sigma_{y}$ term in Eq.~(\ref{WeylHamEq})
are discussed in Sec.~\ref{Weyl_magnetoconductivity}, Sec.~\ref{carrierdensitydep}, and Sec.~\ref{signsigmaz}, respectively.

\subsection{Magnetoconductivity} \label{Weyl_magnetoconductivity}

We would like to shed light on the role and importance of the linear-in-$\lambda$ terms in the conductivity,
which have not been discussed previously.
The linear $\lambda$ terms come from the interplay between angular dependences of the band
Hamiltonian and of the spin-orbit scattering.
To this end in this subsection we first artificially turn off the angular dependence of the spin-orbit scattering
in order to reproduce known results,\cite{Shan12prb}
then turn them back on and analyze their effect.

At first, we turn off the angular dependence of the spin-orbit scattering.
This means that we only consider the angular average of all impurity lines in Fig.~\ref{All_Diagrams}.
In this way, $U_{\bm{k}\bm{k^{\prime}}} G^{\text{R}}_{0,\bm{k}^{\prime}} U_{\bm{k}^{\prime}\bm{k}}
\to |\mathcal{U}|^{2} [ G^{\text{R}}_{0,\bm{k}^{\prime}}
+ \sigma_{z} G^{\text{R}}_{0,\bm{k}^{\prime}} \sigma_{z} \lambda^{2} / 2 ]$
in the Born self-energy and $b_{n\ne0} \to 0$ in the bare vertex.
As a result, the scattering time is $\tau  \to \tau_{0} / [ ( 1+ \lambda^{2}/2 ) ( 1 + b^{2} )]$
and the transport time are
\begin{equation}\label{tautr_angularind}
\tau^{\text{tr}} \to 2 \tau_{0} / [ 1 + 3 b^{2} + (3 + b^{2}) \lambda^{2} / 2],
\end{equation}
which agrees with Ref.~[\onlinecite{Shan12prb}]. However  $1/\tau^{\text{tr}}$, due to angular structure of the Diffuson, is very different between
Ref.~[\onlinecite{Shan12prb}] and our paper. Even more significantly,  Ref.~[\onlinecite{Shan12prb}] misses linear in $\lambda$ terms in both scattering and transport times which as we argue below lead to non-trivial physics.

\begin{figure}[t!]
\begin{center}
\includegraphics[width=1.04\columnwidth]{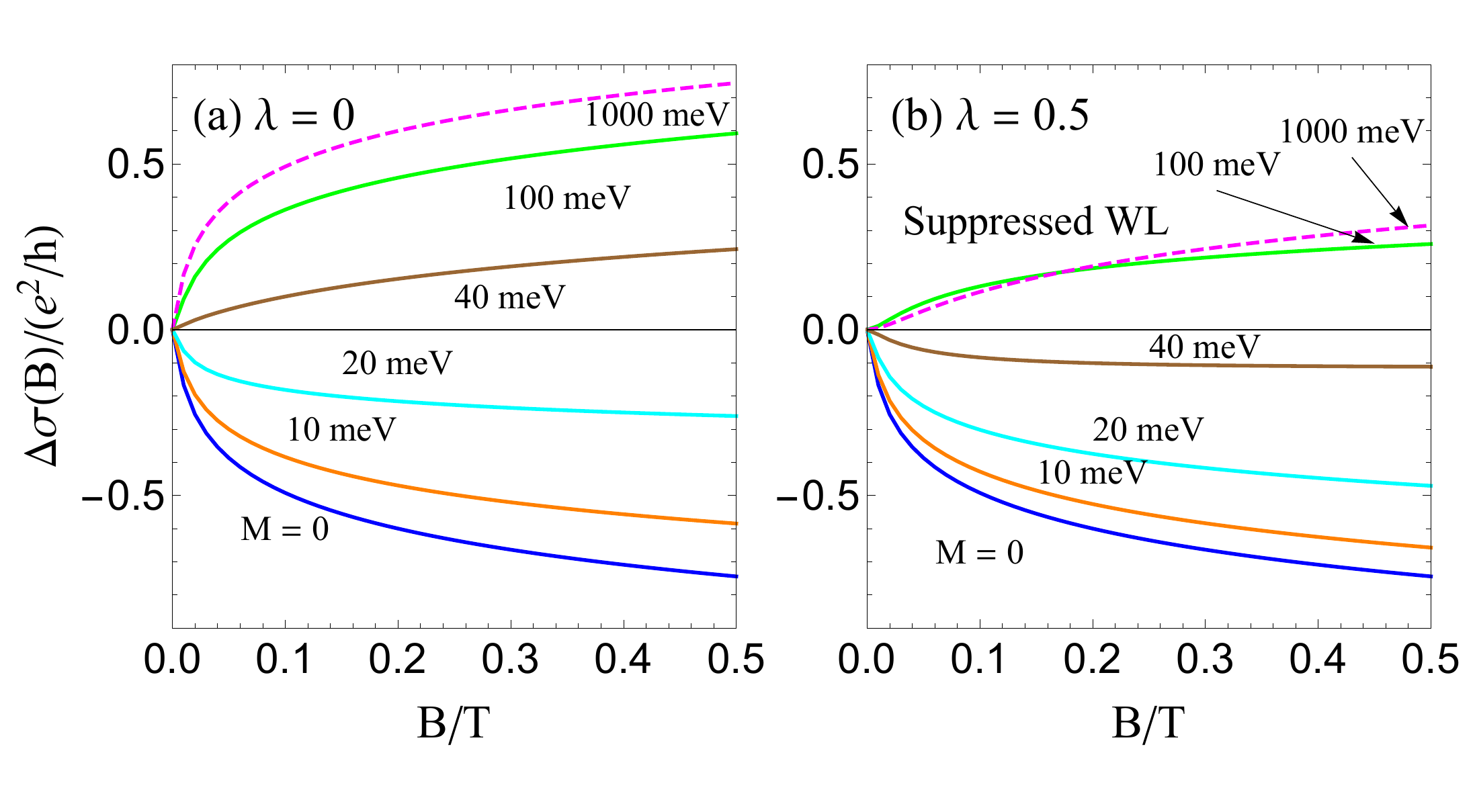}
\caption{(Color online) The magnetoconductivity plots obtained without considering the angular-dependent ($\propto \lambda$) part in the Green's functions.
\textbf{(a)} and \textbf{(b)} magnetoconductivity $\Delta \sigma (B)$ for different masses $M$
at weak spin-orbit scattering $(\lambda = 0)$ and strong spin-orbit scattering ($\lambda = 0.5$).
Comparison between the two panels confirms the suppression of WL due to spin-orbit scattering.
For the effect of the linear-in-$\lambda$ terms, see Fig.~\ref{eta1_Plot}(b).
Parameters are $A = 300$ meV$\cdot$nm, $n_{e} = 0.01\text{ nm}^{-2},$
$n_{\text{i}} = 0.0001\text{ nm}^{-2}$ and $\ell_{\phi} = 500$ nm. Our results by the Keldysh formalism are qualitatively consistent with those by the Kubo formula \cite{Shan12prb}, verifying the validity of our approach.}
\label{eta0_Plot}
\end{center}
\end{figure}

The magnetoconductivity obtained without considering the dependence of the spin-orbit scattering
is plotted in Fig.~\ref{eta0_Plot}.
In Fig.~\ref{eta0_Plot}(a), in the absence of the spin-orbit scattering,
the crossover between the weak antilocalization (WAL) and WL
is displayed when modulating $M$, as featured in Ref.~[\onlinecite{Lu11prl}].
In Fig.~\ref{eta0_Plot}(b), the suppression of weak localization in the presence of the spin-orbit impurities is shown. Moreover, in Fig.~\ref{eta0_Plot}, the blue line that corresponds to the massless limit changes noticeably when the spin-orbit scattering is included. 
In the end, comparison between Figs.~\ref{eta0_Plot}(a) and \ref{eta0_Plot}(b) confirms the suppressed WL due to the spin-independent part of the spin-orbit scattering, as argued in Ref.~[\onlinecite{Shan12prb}].
Noticeably, in this approach, the WAL channel number $(\alpha_{1})$ that emerges in the massless limit is
\begin{equation}\label{alpha1_massless_eta0}
\alpha_{1} \equiv - 1/2.
\end{equation}
The absence of any $\lambda$-dependence in Eq.~(\ref{alpha1_massless_eta0}) does not violates the universality requirement of the symplectic magnetoconductivity: the weak antilocalization channel number is expected to be universal up to the second order in $\lambda$ in the massless limit.
However, this cannot in general justify neglecting the linear-$\lambda$ term.

\begin{figure}[b!]
\begin{center}
\includegraphics[width=0.7\columnwidth]{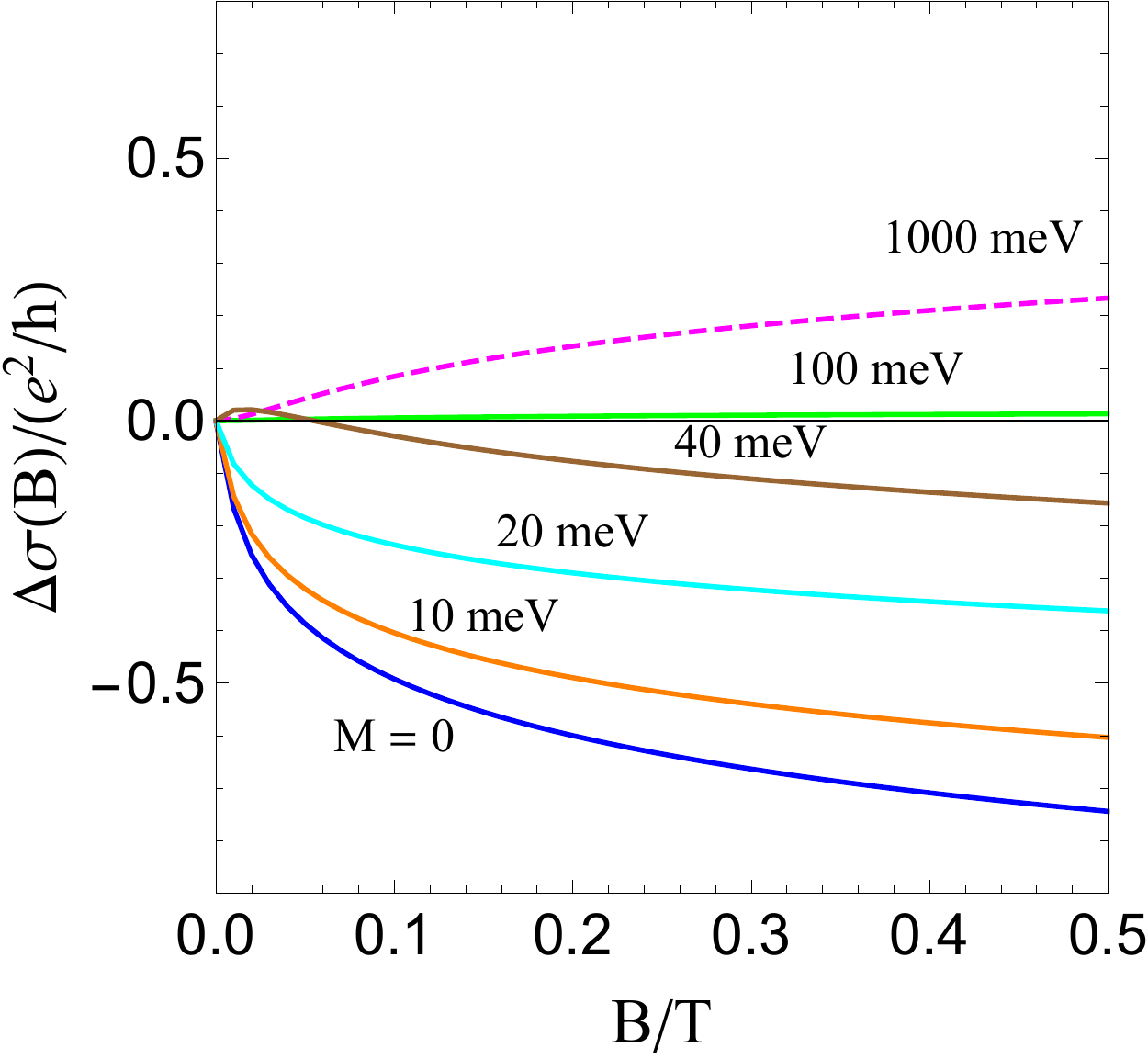}
\caption{(Color online) The magnetoconductivity $\Delta \sigma (B)$ plots at $\lambda = 0.5$ for different masses $M$
with considering the angular-dependent part in the spin-orbit scattering.
Compared with Fig.~\ref{eta0_Plot}(b),  it is shown that more suppression of WL channel when the angular-dependent part
of the spin-orbit scattering is taken into account.}
\label{eta1_Plot}
\end{center}
\end{figure}

Next, we switch on the $\lambda$-linear term in the Green's functions, with the results shown in Fig.~\ref{eta1_Plot}. Comparing Fig.~\ref{eta1_Plot} with Fig.~\ref{eta0_Plot}(b),
a much stronger suppression of the WL channel $(\alpha_{2})$ is seen due to the $\lambda$-linear terms, in particular for large $M$. 
Moreover, the WAL channel number $(\alpha_{1})$ in the massless limit is exactly 1/2, which also satisfies the universal condition protected by time-reversal symmetry.

\subsection{Zero-field conductivity}\label{carrierdensitydep}

At zero magnetic field, the electrical conductivity includes the classical Drude conductivity $\sigma^{\text{Dr}}$ and the quantum-interference conductivity $\sigma^{\text{qi}} (0),$
with $\sigma^{\text{Dr}}$ the dominant term.

\begin{figure}[t!]
\begin{center}
\includegraphics[width=0.8\columnwidth]{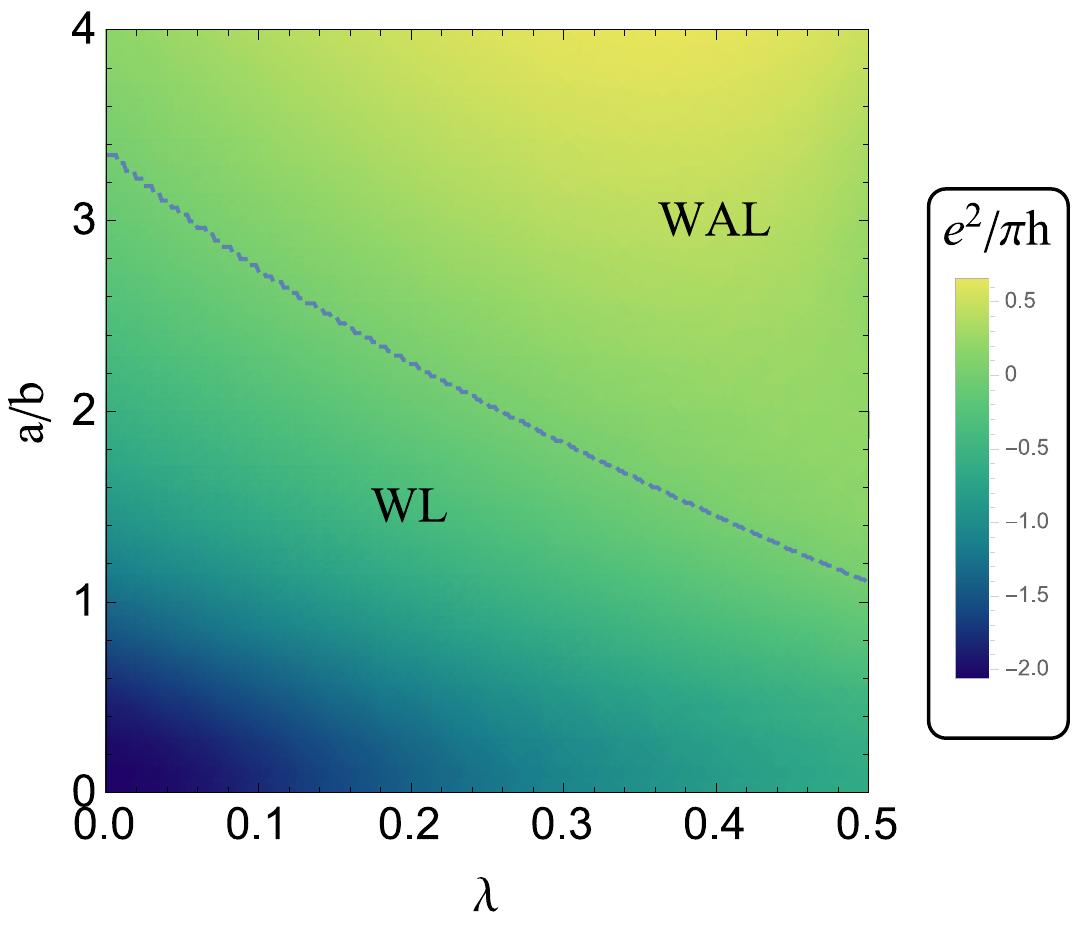}
\caption{(Color online) Zero-field quantum conductivity $\sigma^{\text{qi}}(0)$ in terms of $a/b$ and $\lambda.$
The unit of the conductivity is $e^{2} / h$ with the color bar on the right.
The blue dashed line separates the WAL and WL regimes. The parameters here are the same as in Fig.~\ref{eta0_Plot}.}
\label{WALWLphase}
\end{center}
\end{figure}

\textit{Zero quantum conductivity.}---The zero-field quantum conductivity $\sigma^{\text{qi}} (0)$ can be used to locate the crossover between WAL and WL, and then separate out the regimes in which they occur, as shown in Fig.~\ref{WALWLphase}. In Fig.~\ref{WALWLphase} the $\lambda$-linear terms are taken into account from the start. A positive/negative $\sigma^{\text{qi}} (0)$ corresponds to WAL/WL, respectively. 
From Fig.~\ref{WALWLphase}, in the absence of spin-orbit scattering, the WAL/WL transition occurs at $Ak_F/M =a/b \sim 3.3$. Naively, one would expect that the unitary symmetry point occurs for $A k_F=M$. However, this limit is shifted due to non-trivial coupling between singlet and triplet Cooperon channels.
 As the strength of the spin-orbit scattering is increased, the WAL/WL transition happens at smaller $a/b$. The spin-orbit scattering pushes the WAL/WL boundary, so the WAL regime becomes far broader. On the other hand, the behavior at large mass is similar to the 2D conventional electron gas case where the spin-orbit scattering drives the system from the WL to WAL regimes.

\textit{Carrier density dependence.}---One interesting feature of Ref.~[\onlinecite{Tkachov11prb}] is
the detailed study of the carrier density dependence.
Although those studies were only for the 2D TIs in Ref.~[\onlinecite{Tkachov11prb}],
it motivated authors to investigate the carrier density dependence of the zero-field conductivity.
Note that a back gate is usually applied to adjust the carrier density,
whereas Hall measurements are performed to measure the density.
We introduce $\lambda = \lambda_{c} n_{e}$ since $\lambda \propto k_{\text{F}}^{2}$ gives a linear density dependence.
Note that $\lambda_{c} = 4 \pi \lambda_{0}$ where $\lambda_{0}$ is introduced in Sec.~\ref{sec::ImpurityPotential}.

At zero temperature, the residual conductivity \cite{Rossi12prl} is
$\sigma^{(0)} = \sigma^{\text{Dr}}_{xx} + \sigma^{\text{qi}}_{xx}$, and $\ell_{\phi}$ in Eq.~(\ref{zerofieldcondu}) should
be replaced by the sample size $L,$ because the former will be divergent if $T \to 0$.
By tuning the gate voltage, the carrier density dependence of $\sigma^{(0)}$
can be experimentally measured and the strength
of the spin-orbit scattering can be extracted from $\sigma^{(0)}$. This is one of the central arguments of this work.
To date in the literature only $\lambda^{2}$ terms in $\tau$ or $\tau_{\text{tr}}$ have been identified,
which arise when the spin-orbit scattering terms are averaged over directions in momentum space assuming a circular Fermi surface.
It contributes a negligible $n_{e}^{2}$ dependence to $\sigma^{(0)}$, which is effectively linear in $n_e$ due to Drude part.
When the nontrivial linear-$\lambda$ term is taken into account in $\tau$ and $\tau_{\text{tr}}$,
which is caused by the non-commutativity of the band structure and the random spin-orbit impurity field, the WAL/WL correction has a more pronounced dependence on the carrier number density. This provides a new possibility to extract the spin-orbit scattering constant from the density dependence of $\sigma^{(0)}.$

\begin{figure}[t!]
\begin{center}
\includegraphics[width=0.7\columnwidth]{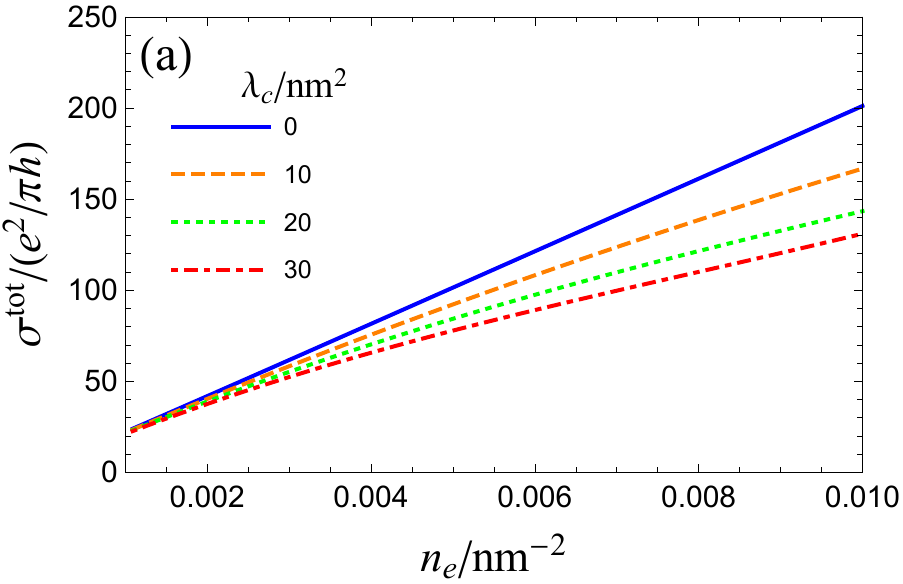} \\[3ex]
\includegraphics[width=0.71\columnwidth]{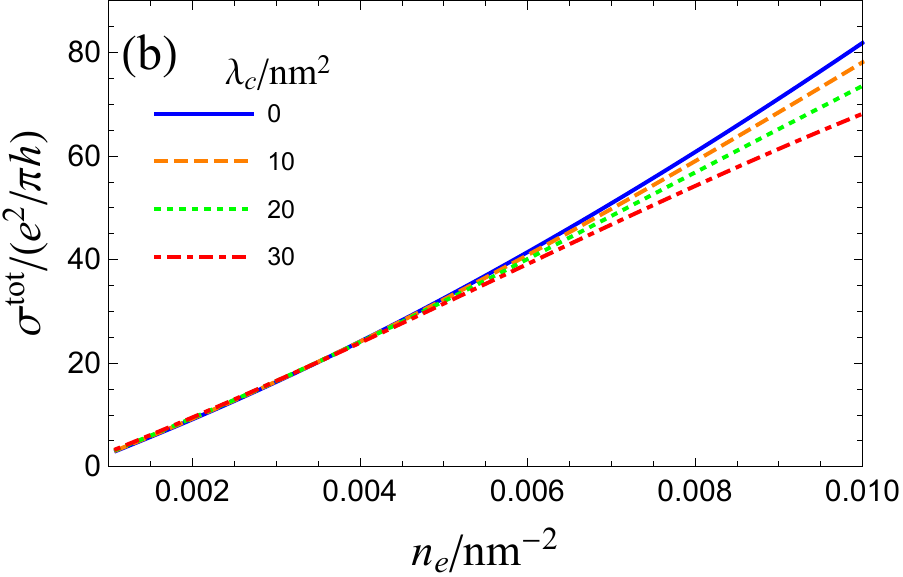}
\caption{(Color online) The carrier density dependence of the total conductivity
$\sigma^{\text{tot}} = \sigma^{(0)}$ at the massless $(M=0)$ (a)
and massive $(M=100\text{ meV})$ (b) limits.
Parameters are the same as Fig.~\ref{eta0_Plot} and $L$ is set to be equal to $\ell_{\phi}$ for simplicity.
Note that $\lambda = \lambda_{c} n_{e}$ and $ A k_{\text{F}} = 106$ meV for $n_{e} = 0.01\,\text{nm}^{-2}.$}
\label{Weyl_density}
\end{center}
\end{figure}

The carrier density dependence of the \textit{total conductivity} for different values of the spin-orbit scattering strength is shown in Fig.~\ref{Weyl_density}. In Fig.~\ref{Weyl_density}(a), the massless limit is considered and the density dependence deviates considerably from the linear form of the Drude contribution (in the approximation use here of short-range impurities). In Fig.~\ref{Weyl_density}(b), the mass term $(M=100\text{ meV})$ suppresses the density dependence due to the extrinsic spin-orbit scattering.

\begin{figure}[t!]
\begin{center}
\includegraphics[width=0.7\columnwidth]{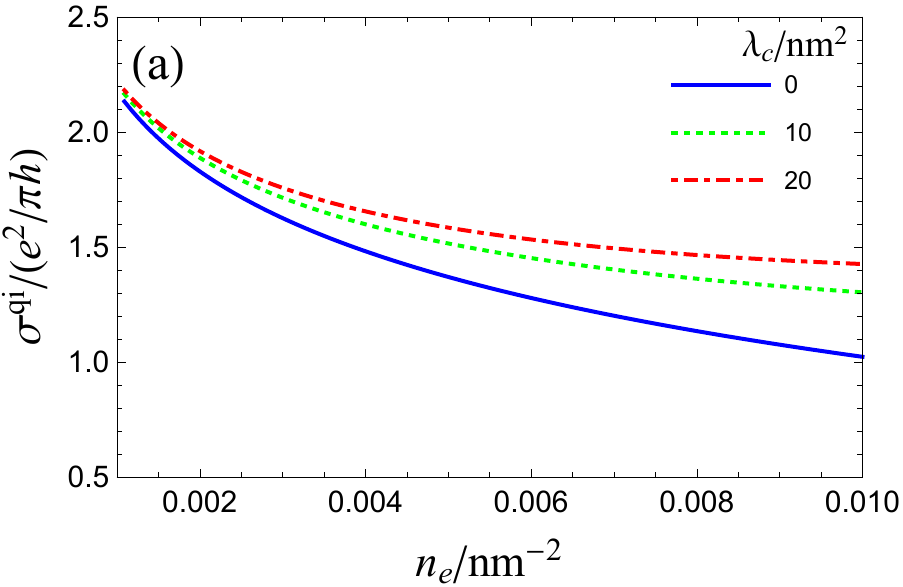}\\
\includegraphics[width=0.7\columnwidth]{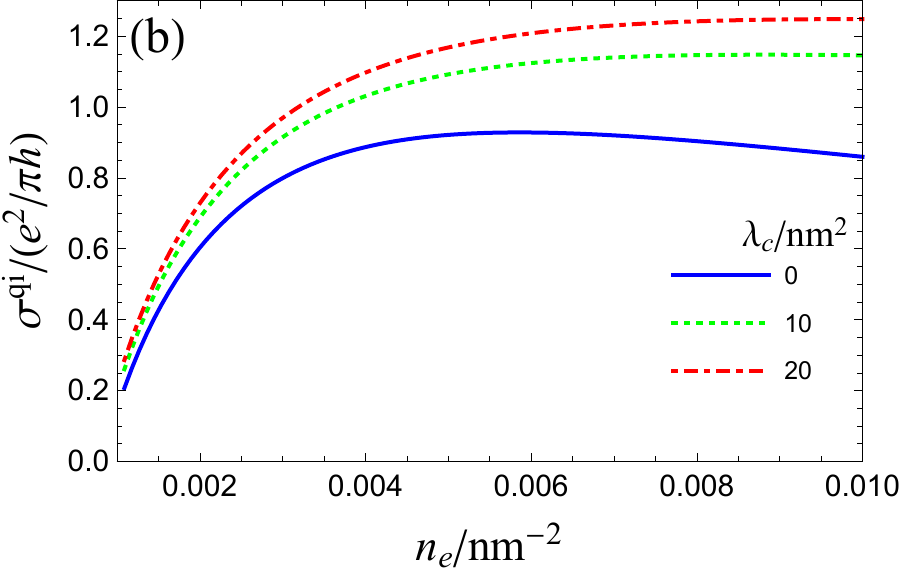}\\
\includegraphics[width=0.72\columnwidth]{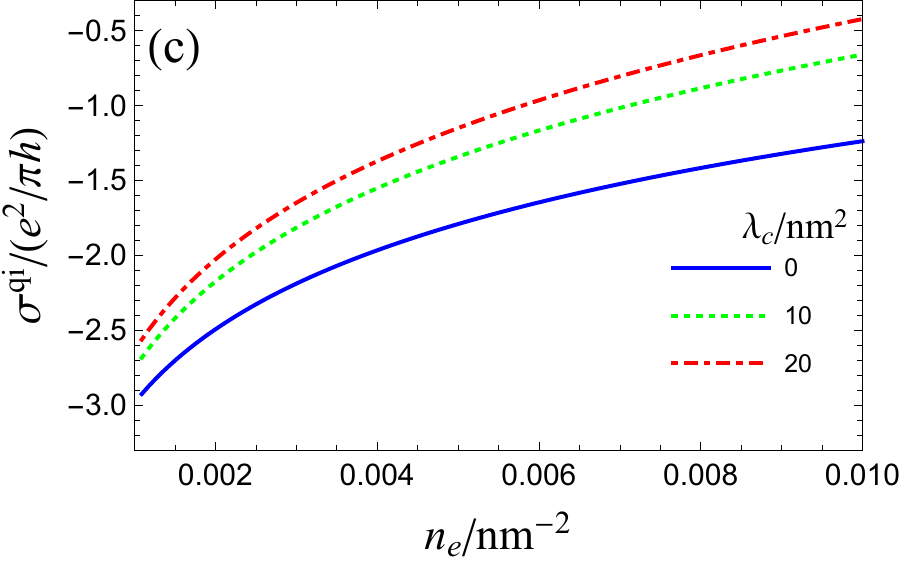}\\
\caption{(Color online) The carrier density dependence
of the quantum-interference conductivity $\sigma^{\text{qi}}$
at the massless $(M = 0)$ (a),  small mass $( M = 10 \text{ meV})$ (b), and large mass $(M = 100 \text{ meV})$ (c) cases.
Parameters here are the same as Fig.~\ref{Weyl_density}.}
\label{Weyl_density_sigmqi}
\end{center}
\end{figure}

In light of this finding, and of the fact that the WL/WAL contribution can be separated from the Drude contribution in an experiment, it is enlightening to plot the carrier density dependence of the quantum-interference part of the conductivity \textit{alone}. The carrier density dependence of the quantum-interference part of the conductivity is plotted in Fig.~\ref{Weyl_density_sigmqi} for three different masses : $M = 0,$ 10 meV, and 100 meV. In the massless limit, Fig.~\ref{Weyl_density_sigmqi}(a), for $\lambda_c = 0$ the contribution $\sigma^{\text{qi}}_{xx}$ follows the logarithmic dependence expected of a 2DEG. As $\lambda_c$ increases this logarithm becomes nearly flat at large enough densities. This altered density dependence arises from the linear terms in the scattering and transport times. In the small mass limit, Fig.~\ref{Weyl_density_sigmqi}(b), a sharp suppression of the conductivity at the small density is displayed, because the effect of the mass term becomes more significant when the carrier density decreases and the WAL channel is suppressed. In the large mass limit, see Fig.~\ref{Weyl_density_sigmqi}(c), a negative conductivity correction (WL) is expected.

\subsection{Sign reversal of the $\sigma_{y}$ term in Eq.~(\ref{WeylHamEq})}\label{signsigmaz}

Ref.~[\onlinecite{Pan15scirep}] suggests an alternative model for describing Dirac fermions~:
\begin{equation}\label{Dirac_Ham}
H_{0\bm{k}} = A (\sigma_{x} k_{x} - \sigma_{y} k_{y}) + M \sigma_{z}.
\end{equation}
Compared with Eq.~(\ref{WeylHamEq}), Eq.~(\ref{Dirac_Ham})
gives the same energy dispersion but with an opposite sign on the $k_{y}$ momentum,
which means a chirality reversal.
Since the angular dependence of the spin-orbit scattering is taken into account, the change on the
angular-dependence of the band Hamiltonian $H_{0\bm{k}}$ will also affect the results in Secs.~\ref{Weyl_magnetoconductivity}
and~\ref{carrierdensitydep}.
In this subsection, we will show the effect of the sign reversal of the $\sigma_{y}$ term,
and with a comparison between two cases, the effect of the angular-dependent part of the spin-orbit scattering
might become more clear.

For the model described by Eq.~(\ref{Dirac_Ham}), the scattering time is
\begin{equation}\label{Yang_tau_SO}
\tau = \tau_{0} / \big[ ( 1+ \lambda^{2} / 2)(1 + b^{2} ) - \lambda a^{2} \big],
\end{equation}
which matches Ref.~[\onlinecite{Adroguer15prb}] in the massless limit
and has an opposite sign in the linear $\lambda$ term when comparing with Eq.~(\ref{tau}),
and the transport time becomes
\begin{equation}\label{Yang_trsptime}
\tau_{\text{tr}}  = 2 \tau_{0}/ \big[ 1 + 3 b^{2} - 2 a^{2} \lambda +
\big( 5 + 3 b^{2}\big) \lambda^{2} /4 \big],
\end{equation}
except an opposite sign in the linear $\lambda$ term from Eq.~(\ref{trsptime}).
These opposite signs of linear $\lambda$ terms come from the opposite sign of the $\sigma_{y}$ term
in the band Hamiltonian,
which implies that linear $\lambda$ terms in $\tau$ and $\tau_{\text{tr}}$ are
due to the interplay between the band Hamiltonian and the spin-orbit scattering.

Furthermore, the WAL channel number $(\alpha_{1})$ remains the same,
while the rest channel numbers $(i = 2, 3)$ become $\alpha_{i} = \alpha_{i,j} (- \lambda)^{j}$ and
the effective coherence length coefficients become $\gamma_{i} = \gamma_{i,j} (-\lambda)^{j},$
where $\alpha_{i,j}$ and $\gamma_{i,j}$ are listed in Tab.~\ref{alphagammaexp}.
Since $j \le 2,$ the changes only happens on the linear $\lambda$ term.
The unchanged WAL channel number indicates that the sign reversal of the $\sigma_{y}$ term
does not affect the WAL behavior, because there is no net spin in the singlet channel.
The $\lambda \to -\lambda$ phenomena of triplet channel numbers under the sign reversal of the $\sigma_{y}$ term
is expected because of the net spin of triplet channels.

\section{Discussion}
\label{Discussions}

The focus of our work is to analyze the strong effect of terms linear in the spin-orbit scattering strength on weak localization and antilocalization in WSM thin films. In contrast to previous works, \cite{Shan12prb} where only the angle-averaged (second-order in $\lambda$) contribution is considered in the scattering time $\tau$, we have demonstrated that a correction to first order in the spin-orbit scattering strength is present, which has a strong dependence on the relative angle between the incoming and outgoing wave vectors. It emerges from the angular integration of the chiral Dirac fermion Hamiltonian and the disorder Hamiltonian describing spin-orbit scattering. This correction is one order of magnitude larger than that found in Ref.~[\onlinecite{Shan12prb}], making it a significant perturbation.

In the presence of strong spin-orbit scattering, the suppression of the WL channel is expected on general grounds since this channel corresponds to one of the spin-polarized triplet channels. Spin-orbit scattering randomizes the spin polarization in the triplet channels and eventually suppresses the contribution due to these. The suppression is already seen even when only the isotropic part of the spin-orbit scattering ($\propto \lambda^2$) is taken into account, as in Ref.~[\onlinecite{Shan12prb}]. The $\lambda$-linear terms enhance this suppression, as shown in Fig.~\ref{eta1_Plot}.

An interesting qualitative change due to the extrinsic spin-orbit scattering is observed in the transition from WL to WAL occurring as a function of the mass. This may be seen in Fig.~\ref{WALWLphase}. At small and large values of the mass, as expected, the extrinsic spin-orbit scattering has little effect on the WL/WAL correction. At large mass, moreover, the extrinsic spin-orbit scattering also plays a negligible role in the momentum relaxation time. However, when the system is close to the unitary symmetry class, the extrinsic spin-orbit scattering is critical in determining whether the system experiences WL or WAL.
A rough estimate of WL/WAL transition line in Fig.~\ref{WALWLphase} is $a/b \sim 1/(0.3 + 0.8 \cdot \lambda),$
which exists when $n_{e} \in [0.01, 0.02] \, \text{nm}^{-2}$ and $\ell_{\phi} \in [200,500] \, \text{nm}.$
This fitting equation can provide a semi-empirical formula to extract the $\lambda,$
and the extraction of $\lambda$ can be examined by another method as introduced in the following.

At small mass, the extrinsic spin-orbit scattering makes an important contribution to the momentum relaxation time, causing it to acquire a strong density dependence. This manifests itself in a flattening of the 2D WAL correction as a function of carrier number density. To extract the spin-orbit scattering strength $\lambda$ from $\sigma^{\text{qi}} (0)$ in the massless limit, it is possible to follow the fitting equation
\begin{equation}\label{fittingequation}
\sigma^{\text{qi}} (0) = a_{0} \ln n_{e} + b_{0} n_{e}.
\end{equation}
Here we assume zero temperature, so $\ell_{\phi}$ is replaced by the sample size $L$ as mentioned above. The extracted coefficient $b_{0}$ yields the spin-orbit scattering strength
$\lambda =  - b_{0} n_{e} / (3 e^{2} / 2 \pi h)$ when $\lambda \ll 1$. The short-range impurity potential that we have used in this article needs to be replaced by a long-range Coulomb potential (or a generic long-range potential). In the latter case, the Bethe-Salpeter equation (\ref{BetheSalpeter}) will involve an additional angular integration over the impurity potential, in which case it is no longer solvable in closed form. One possible rough estimate for $\sigma^{\text{qi}} (0)$ can be obtained by retaining the form of Eq.~(\ref{zerofieldcondu}) while substituting the same $\tau$ for the long-range potential as was found above for the short-range case. In the long-range case, the normal impurity potential $\mathcal{U}$ for 2D massless Dirac fermions becomes $\mathcal{U}_{\text{long}} = Z e^{2} / [2 \varepsilon_{r} k_{\text{F}} \sin^{2} (\gamma / 2)],$\cite{Culcer08prb} where $\varepsilon_{r}$ is the material-specific dielectric constant. Thus, the long-range potential $\mathcal{U}_{\text{long}}$ will have the same $n_{e}$ dependence as the short-range case, and the fitting equation~(\ref{fittingequation}) will be still correct.

\section{Conclusions}\label{Conclusions}

In conclusion, we have studied theoretically the weak localization/antilocalization correction to the conductivity of ultra-thin films of Weyl semimetals. We have considered scalar and spin-orbit scattering mechanisms on the same footing. The non-commutativity of the matrix Green's functions and spin-dependent impurity potential gives rise to terms linear in the extrinsic spin-orbit scattering strength which play an important role in determining whether the system experiences WL or WAL and strongly affect the density dependence of the WAL correction in the massless limit. In addition, these terms restore the universality of the WAL channel number in the massless limit, which appears to be violated when only the second-order terms are considered.

\acknowledgments

This research was supported by the Australian Research Council Centre of Excellence in Future Low-Energy Electronics Technologies (project number CE170100039) and funded by the Australian Government, NBRPC (Grant No.~2015CB921102), the National Key R \& D Program (Grant No.~2016YFA0301700), and the National Natural Science Foundation of China (Grants No.~11534001, No.~11474005 and No.~11574127). E. M. H. acknowledges financial support by the German Science Foundation (DFG) via SFB 1170 "ToCoTronics" and the ENB Graduate School on Topological Insulators.

\bibliographystyle{apsrev4-1}
\bibliography{Weyl_WAL_SO} 

\end{document}